\documentclass[preprint]{aastex62}
\usepackage{float}

\usepackage{appendix}
\usepackage{mathtools}
\usepackage{epsfig}
\usepackage{array}
\usepackage{graphicx}
\usepackage{graphics}
\usepackage{subfigure}
\usepackage[latin1]{inputenc}
\usepackage{latexsym}

\usepackage{threeparttable}

\shorttitle{Early SNe~Ia} \shortauthors{Li et al.}

\def\gsim{\;\lower4pt\hbox{${\buildrel\displaystyle >\over\sim}$}\;}
\def\lsim{\;\lower4pt\hbox{${\buildrel\displaystyle <\over\sim}$}\;}
\def\grls{\;\lower4pt\hbox{${\buildrel\displaystyle >\over <}$}\;}
\begin{document}

\title{Can Helium-detonation Model Explain the Observed Diversity of Type Ia Supernovae?}
\correspondingauthor{Xiaofeng Wang, Wenxiong Li}
\email{wang\_xf@mail.tsinghua.edu.cn, li-wx15@tsinghua.org.cn}
%Holmbo, S 13gy data
%Peter Brown 2017erp data
%2017cfd Weikang Zheng
%2009ig Jozsef Vinko
%19ein 1st spectrum ask andy?
%latetime Fe velocity Kate
%2019ein nebular velocity
%uniform luminosity for sub-ch model
\author{Wenxiong Li}
\affil{Physics Department and Tsinghua Center for Astrophysics (THCA), Tsinghua University, Beijing, 100084, China}
\affil{The School of Physics and Astronomy, Tel Aviv University, Tel Aviv 69978, Israel}
\author{Xiaofeng Wang}
\affil{Physics Department and Tsinghua Center for Astrophysics (THCA), Tsinghua University, Beijing, 100084, China}
\affil{Beijing Planetarium, Beijing Academy of Science and Technology, Beijing, 100089}
\affil{Purple Mountain Observatory, Chinese Academy of Sciences, Nanjing, 210023, China}
\author{Mattia Bulla}
\affil{Nordita, KTH Royal Institute of Technology and Stockholm University, Roslagstullsbacken 23, 106 91 Stockholm, Sweden}
\author{Yen-Chen Pan}
\affil{Graduate Institute of Astronomy, National Central University, 300 Jhongda Road, Zhongli, Taoyuan, 32001, Taiwan}
\author{Lifan Wang}
\affiliation{George P. and Cynthia Woods Mitchell Institute for Fundamental Physics \& Astronomy, \\
Texas A. \& M. University, Department of Physics and Astronomy, 4242 TAMU, College Station, TX 77843, USA}
%\affiliation{Purple Mountain Observatory, Nanjing 210023, People's Republic of China}
\author{Jun Mo}
\affil{Physics Department and Tsinghua Center for Astrophysics (THCA), Tsinghua University, Beijing, 100084, China}
\author{Jujia Zhang}
\affil{Yunnan Observatories (YNAO), Chinese Academy of Sciences (CAS), Kunming, 650216, China}
\affil{Key Laboratory for the Structure and Evolution of Celestial Objects,CAS, Kunming, 650216, China}
\author{Chengyuan Wu}
\affil{Physics Department and Tsinghua Center for Astrophysics (THCA), Tsinghua University, Beijing, 100084, China}
\author{Jicheng Zhang}
\affil{Physics Department and Tsinghua Center for Astrophysics (THCA), Tsinghua University, Beijing, 100084, China}
\author{Tianmeng Zhang}
\affil{Key Laboratory of Optical Astronomy, National Astronomical Observatories, Chinese Academy of Sciences, 10101, Beijing}
\affil{School of Astronomy and Space Science, University of Chinese Academy of Sciences, 101408, Beijing}
\author{Danfeng Xiang}
\affil{Physics Department and Tsinghua Center for Astrophysics (THCA), Tsinghua University, Beijing, 100084, China}
\author{Han Lin}
\affil{Physics Department and Tsinghua Center for Astrophysics (THCA), Tsinghua University, Beijing, 100084, China}
\author{Hanna Sai}
\affil{Physics Department and Tsinghua Center for Astrophysics (THCA), Tsinghua University, Beijing, 100084, China}
\author{Xinghan Zhang}
\affil{Physics Department and Tsinghua Center for Astrophysics (THCA), Tsinghua University, Beijing, 100084, China}
\author{Zhihao Chen}
\affil{Physics Department and Tsinghua Center for Astrophysics (THCA), Tsinghua University, Beijing, 100084, China}
\author{Shengyu Yan}
\affil{Physics Department and Tsinghua Center for Astrophysics (THCA), Tsinghua University, Beijing, 100084, China}
\begin{abstract}
We study a sample of 16 Type Ia supernovae (SNe Ia) having % including four high-velocity (HV) group objects, 
 both spectroscopic and photometric observations within 2 $-$ 3 days after the first light. %SN~2014J is located at the boundary of HV and NV, and the first observation is late. 
 %, 
 The early $B-V$ colors of such a sample      tends to show a continuous distribution. %, in contrast to the previous claim for a bimodal trend.
%The early color dispersion is $\sim 0.6$ mag and gradually decreases to $\sim 0.3$ mag.
For objects with normal ejecta velocity (NV), the C~II $\lambda$6580 feature is always visible in the early spectra while it is absent or very weak in the high-velocity (HV) counterpart. 
Moreover, the velocities of the detached high-velocity features (HVFs) of Ca~II NIR triplet (CaIR3) above the photosphere are found to be much higher in HV objects than in NV objects, with typical values exceeding 30,000 km~s$^{-1}$ at 2 $-$ 3 days. %The He-detonation model can naturally produce AE effect of Si and Ca in the direction of the northern hemisphere (     n}$_1$-     n}$_3$ in Figure \ref{fig:carbon}) through the first He detonation, and produce HVFs with extremely high velocities observed in the spectra. 
We further analyze the relation between %velocities of Si~II~$\lambda$6355 at maximum, $v_{\rm Si,max}$,
the velocity shift of late-time [Fe~II] lines ($v_{\rm [Fe~II]}$) and host galaxy mass. We find that all HV objects have redshifted $v_{\rm [Fe~II]}$ 
%and are located in massive galaxies 
while NV objects have both blue- and redshifted $v_{\rm [Fe~II]}$.
%and are located in both massive and low-mass galaxies. %SNe with redshifted $v_{\rm [Fe~II]}$ require an asymmetric explosion model such as He-detonation model. 
It is interesting to point out that the objects with redshifted $v_{\rm [Fe~II]}$ are all located in massive galaxies, implying that HV and a portion of NV objects may have similar progenitor metallicities and explosion mechanisms. 
%Combined these features with numerical simulation results, 
We propose that, with a geometric/projected effect, the He-detonation model may account for the similarity in birthplace environment and the differences seen in some SNe Ia, including $B-V$ colors, C~II feature, CaIR3 HVFs at early time and $v_{\rm [Fe~II]}$ in the nebular phase.
%among them might be caused by different viewing angles, which is consistent with previous study that some NV objects may have the same origin as HV group.
Nevertheless, some features predicted by He-detonation simulation, such as the rapidly decreasing light curve, deviate from the observations, and some NV objects with blueshifted nebular $v_{\rm [Fe~II]}$ may involve other explosion mechanisms. %More sample with early color and spectroscopic information is needed to test different explosion models. %, . % obtained by numerical simulation. 
%More evidence is required to determine whether %HV objects are different SNe from NV objects or 
%some NV objects and HV objects have a common origin. % but different viewing angles. %, and not all NV can be explained by this scenario. %
%Nonetheless, our work confirms that HV group is a distinctive subgroup of SNe Ia.
\end{abstract}

\section{Introduction}
Type Ia supernovae (SNe Ia) are thought to be the thermonuclear runaway of a carbon-oxygen (C/O) white dwarf (WD) in a binary system. However, the physical properties of the companion star and the explosion mechanism of normal SNe Ia are still unclear \citep{2011NatCo...2..350H}. One mechanism is the delayed-detonation (DDT) model: ignition of a Chandrasekhar-mass ($M\rm_{Ch}$) WD, which requires expansion before the explosion \citep{1984ApJ...286..644N,1991AA...245..114K}. In this scenario, a WD accretes material from a non-degenerate companion star until it reaches $M\rm_{Ch}$ and explode. Several three-dimensional simulations of near M$_{Ch}$ explosions have been published \citep{2005ApJ...623..337G,2007AA...464..683R,2008AA...478..843B,2011MNRAS.414.2709S,2012ApJ...750L..19R,2013MNRAS.429.1156S}. Light curves and spectra of a normal SN~Ia can be produced by this scenario \citep{2013MNRAS.429.1156S,2013MNRAS.436..333S}.      \cite{2015MNRAS.448.2766B} suggest the DDT model succeeds in reproducing observables of high-velocity subgroup (see below). Another mechanism for normal SNe Ia is the violent merger of two WDs \citep{1984ApJS...54..335I,1984ApJ...277..355W}. Radiative transfer simulations for relatively massive WD pairs have
shown reasonable match with spectra and light curves of normal SNe Ia \citep{2012ApJ...747L..10P,2014ApJ...785..105M}. However, the strong polarisation features arising from the significant asymmetries in this scenario are not observed in most SNe Ia \citep{2016MNRAS.455.1060B}.

Another explosion mechanism for SNe Ia is the sub-$M\rm_{Ch}$ He-detonation (or double detonation) model. There are two detonation process in this model: the first one is the detonation of the accreted He on the surface of the progenitor, then the supersonic wave propagates on the surface, converges to the opposite side and trigger the second one \citep{1982ApJ...253..798N,1982ApJ...257..780N,1990ApJ...354L..53L,1994ApJ...423..371W,1996ApJ...472L..81H,2011ApJ...734...38W}. However, earlier-time models needed a thick He shell ($\sim 0.1$ M$_\odot$) and produced a Ti and Ni-rich outer layer which is not observed in SNe Ia. %Both SD and DD scenario can have He-detonation process. In the former, the He layer can be formed through the accretion from a He star \citep{2011ApJ...734...38W}. In the latter, the primary WD can accrete He from a He WD \citep{2010ApJ...709L..64G} or a CO WD with a He layer \citep{2013ApJ...770L...8P}. 
Recent studies reveal that the minimal He-shell mass required to trigger a runaway explosion can be much lower than previous estimates \citep{2014ApJ...797...46S,2019ApJ...878L..38T} and the results of simulations match the main observable of SNe Ia \citep{2010ApJ...719.1067K,2019ApJ...878L..38T,2019ApJ...873...84P}.

Normal SNe Ia can be divided into two categories based on the Si~II~$\lambda$6355 velocities at the maximum ($v_{\rm Si,max}$): high-velocity (HV) ($v_{\rm Si,max}$ $>$ 11,800 km~s$^{-1}$) and normal-velocity (NV) ($v_{\rm Si,max}$ $\leq$ 11,800 km~s$^{-1}$ group \citep{2009ApJ...699L.139W}. \cite{2009ApJ...699L.139W} further found that the HV group is redder and have lower extinction ratio $R_V=A_V/(A_B-A_V)$, where $A_X$ is the extinction in the $X$ band. %than NV group.
\cite{2013Sci...340..170W} found that the $v_{\rm Si,max}$ of SNe Ia show a bimodal distribution, where HV group tends to be located in the inner regions of the host galaxies while the NV group has wider distributions, suggesting different metallicity environments for these two groups. \cite{2015MNRAS.446..354P} and \cite{2020ApJ...895L...5P} supported this idea by showing that HV objects tend to reside in massive galaxies with higher metallicity, while NV objects reside in both low-mass and massive host galaxies. Recently, \cite{2019ApJ...882..120W} found that the HV group has excess blue flux during the early nebular phase, which may be caused by light scattering by the circumstellar material (CSM) around them while NV group does not have such features.

One promising way to study SNe Ia is to observe them as early as possible. Several scenarios only leave detectable imprints in the first few days after the onset of SN explosions. For instance, interaction between the supernova~ejecta and a non-degenerate companion may produce excess flux in the blue optical or ultraviolet (UV) bands %, depending on the binary separation and the viewing angle
\citep{2010ApJ...708.1025K}. %IPTF14atg %a SN~2002es-like subluminous supernova, 
%is the first SN~Ia with strong early UV flux which might originates from the companion interaction  \citep{2015Natur.521..328C}, although \cite{2016MNRAS.459.4428K} found a violent merger of two WDs matches the spectral evolution very well. %The early blue bump of a normal SN~Ia SN~2002cg was consistent with companion interaction scenario, too. However, \cite{2018ApJ...855....6S} ruled out a non-degenerate companion with X-ray and late-time optical observations. 
%The high-cadence,
The very early observations of SN~2017cbv reveal a blue bump, which may favor a subgiant companion (\citealt{2017ApJ...845L..11H};  but see \citealt{2018ApJ...863...24S}). %However,\cite{2018ApJ...863...24S} challenge this interpretation with strict  H/He mass limits derived from a late-time spectrum.
%Recently, SN~2018oh, a normal SN~Ia with long-lasting carbon absorption features \citep{2019ApJ...870...12L}, was discovered in the $Kepler~Space~Telescope$ field with 30-min cadence $Kepler$ photometry since the explosion. \cite{2019ApJ...870L...1D} slightly favor the companion interaction scenario. However, \cite{2019ApJ...870...13S} interpreted the early light curve with shallow $^{56}$Ni distribution. By far, there is no decisive proof for the companion interaction scenario for individual SN~Ia. He-detonation model also predict specific early features with two layers of radioactive elements, which may cause blue-red-blue color evolution at early time \citep{2018ApJ...861...78M,2019ApJ...873...84P}. MUSSES1604D and SN~2018byg are the first two SNe Ia that may have a massive He-shell \citep{2017Natur.550...80J,2019ApJ...873L..18D}. Two more SNe Ia, SN~2016hnk (\citealt{2020ApJ...896..165J}; but see \citealt{2019A&A...630A..76G} for a different interpretation) and SN~2019yvq \citep{2020ApJ...898...56M}, have recently been considered as possible candidates of He-detonation scenario. \cite{2018ApJ...854...52S} claim that the first detonation only produce intermediate mass elements such as Si and Ca that might explain the high-velocity features (HVFs) among SNe Ia at early phase \citep{2005ApJ...623L..37M,2015ApJS..220...20Z,2016ApJ...826..211Z}.

Thanks to the rapid developments of wide-field transient surveys, astronomers have discovered much more young supernovae than in the past, which makes it possible to perform statistical researches of very young SNe Ia. Previous studies mainly focused on photometric properties \citep{2018ApJ...864L..35S,2018ApJ...865..149J,2020ApJ...902...48B,2020ApJ...902...47M}. Here we report the first statistical research that combines very early photometric and spectroscopic properties of normal SNe Ia. In Section \ref{sec:sample} we introduce our sample selection process and observations. In Section \ref{sec:analysis} we analyze several photometric and spectroscopic properties of our sample. In Section \ref{dis} we discuss implications of our results for the explosion mechanism. The discussion and conclusions are presented in Section \ref{con}.

\section{Sample Selection and Observations}\label{sec:sample}
We aim to select SNe Ia with good early multi-band photometric and spectroscopic observations. Peculiar objects, i.e., super-Chandrasekhar SNe Ia, SNe Iax, are not included in our investigation due to possible different physical origins. We focus on SNe Ia with photometry and spectra from $\sim$2 weeks before %to around 
$B$-band maximum, and at least four spectra before the maximum for each object. The sample contains 16 SNe, and the data mainly comes from literature. The basic information  %collected from literature
are listed in Table \ref{info}, including absolute $B$-band peak magnitude, $\Delta m_{\rm 15}(B)$,  $v_{\rm Si,max}$ and sub-type. Among the sample, SN~2017cfd has only two spectra, however, its first spectra is early (at +3.5 days after the first light) and it has dense early photometry, thus we include it for comparison.
%late spectroscopic      and photometric }follow up from $-$10.4 days while .  
%We include SN~2014J for comparison, since it is well-observed after discovery, although it was discovered late.

Besides the data published in the literature, we have some new data for several objects: SNe~2019np, 2019ein (Xi et al. 2020 in prep), 2018yu, 2018gv (Li et al. 2020 in prep), and 2012cg. The broadband $BV$- and Sloan $gri$-band images were obtained with the 0.8~m Tsinghua-NAOC Telescope (TNT) in China  \citep{2012RAA....12.1585H} and processed using the $Zrutyphot$ (Mo et al. in prep.), including bias subtraction and flat fielding.      Template subtraction is performed when reducing these images. The $BV$- and $gri$-band photometry are calibrated into Vega \citep{1992AJ....104..340L} and AB magnitude systems, respectively. The spectra were obtained with the Xinglong 2.16-m telescope (+BFOSC; \cite{2016PASP..128j5004Z}), the Lijiang 2.4-m telescope  \citep[+YFOSC;][]{2019RAA....19..149W}, and the 3.5-m Astrophysics Research Consortium (ARC) telescope. All spectra were reduced using standard IRAF routines.
\startlongtable
\begin{deluxetable}{cccccc}
%\tablecolumns{4} \tablewidth{0pc} 
\tabletypesize{\scriptsize}
\tablecaption{Sample of SNe~Ia with Very Early-time Observations \label{info} }
\tablehead{ \colhead{Name} &\colhead{$M_B$} & \colhead{$\Delta m_{\rm 15}(B)$} & \colhead{$v_{\rm Si,max}$ }   & \colhead{Ref.} & \colhead{Type}\\\colhead{} &\colhead{mag} &\colhead{mag} &\colhead{km~s$^{-1}$} &\colhead{} &\colhead{} }
\startdata
SN 2002bo	&	$-$19.4 $\pm$ 0.4	&	1.13 $\pm$ 0.05	&	13000 $\pm$ 300	&	(1)	& HV\\
SN 2009ig	&	$-$19.46 $\pm$ 0.08	&	0.90 $\pm$ 0.07	&	13500 $\pm$ 200	&	(2)	& HV\\
SN 2011fe	&	$-$19.23 $\pm$ 0.09	&	1.07 $\pm$ 0.06	&	10500 $\pm$ 200	&	(3)	& NV\\
SN 2012cg	&	$-$19.73 $\pm$ 0.02	&	0.86 $\pm$ 0.02	&	10500 $\pm$ 200	&	(4)	& NV\\
SN 2012fr	&	$-$19.49 $\pm$ 0.06	&	0.85 $\pm$ 0.05	&	12120 $\pm$ 70	&	(5)	& HV\\
iPTF13ebh	&	$-$18.95 $\pm$ 0.07	&	1.76 $\pm$ 0.02	&	10400 $\pm$ 200	&	(6)	& NV\\
SN 2013dy	&	$-$19.65 $\pm$ 0.40	&	0.90 $\pm$ 0.03	&	10200 $\pm$ 300	&	(7)	& NV\\
SN 2013gy	&	$-$19.3 $\pm$ 0.2	&	1.23 $\pm$ 0.06	&	10180 $\pm$ 200	&	(8)	& NV\\
iPTF16abc	&	$-$19.56 $\pm$ 0.08	&	0.95 $\pm$ 0.01	&	10200 $\pm$ 300	&	(9)	& NV\\
SN 2017cbv	&	$-$20.04 $\pm$ 0.09	&	1.06 $\pm$ 0.05	&	9600 $\pm$ 400	&	(10)& NV\\
SN 2017cfd	&	$-$19.3 $\pm$ 0.2	&	1.16 $\pm$ 0.11	&	11200 $\pm$ 200	&	(11)& NV\\
SN 2017erp	&	$-$19.1 $\pm$ 0.1	&	1.05 $\pm$ 0.06	&	10400 $\pm$ 200	&	(12)& NV\\
SN 2018gv	&	$-$19.0 $\pm$ 0.3	&	0.96 $\pm$ 0.04	&	11400 $\pm$ 200	&	(13)& NV\\
SN 2018yu	&	$-$19.5 $\pm$ 0.4	&	0.98 $\pm$ 0.01	&	10000 $\pm$ 300	&	(14)& NV\\
SN 2019ein	&	$-$18.81 $\pm$  0.06	&	1.40 $\pm$ 0.01	&	13700 $\pm$ 300	&	(15)& HV	\\
SN 2019np	&	$-$19.1 $\pm$ 0.3	&	0.95 $\pm$ 0.01	&	10000 $\pm$ 100	&	(16)& NV\\
\enddata
\begin{tablenotes}
      \small
      \item Notes. (1) \cite{2004MNRAS.348..261B,2017ApJ...848...66Z}; (2) \cite{2012ApJ...744...38F,2013ApJ...777...40M,2020MNRAS.492.4325S}; (3) \cite{2011Natur.480..344N,2012ApJ...752L..26P}; (4) \cite{2012ApJ...756L...7S,2016ApJ...820...92M} and this work; (5) \cite{2013ApJ...770...29C,2014AJ....148....1Z,2018ApJ...859...24C}; (6) \cite{2015AA...578A...9H}; (7) \cite{2013ApJ...778L..15Z,2016AJ....151..125Z}; (8) \cite{2019AA...627A.174H}; (9) \cite{2018ApJ...852..100M}; (10) \cite{2017ApJ...845L..11H}; (11) \cite{2020ApJ...892..142H}; (12) \cite{2019ApJ...877..152B}; (13) \cite{2020ApJ...902...46Y} and this work; (14) This work; (15) \cite{2020ApJ...893..143K,2020ApJ...897..159P} and this work; (16) This work.
\end{tablenotes}
\end{deluxetable}

\section{Analysis}\label{sec:analysis}
Here we investigate the photometric, spectroscopic and host galaxy properties of our sample. Since the rise times of the light curves are different, we use the time relative to the first light as time coordinate instead of the time relative to the $B$-band maximum, which is commonly used\footnote{We notice that the first-light times from the literature are calculated with different fitting methods and may cause additional uncertainties.}. The first light time is different from the explosion time because there might be a dark phase that lasts for 1 $-$ 2 days \citep{2013ApJ...769...67P}. We collect the first light time of our sample from the literature while calculating the first light time using Lmfit \citep{2016ascl.soft06014N} for SN~2018yu and SN~2019np, yielding a rise time in the $V$ band of 18.6 $\pm$ 0.6 days for SN~2018yu and 19.7 $\pm$ 0.6 days for SN~2019np, respectively. The multi-band light curves of these two SNe are both well-fit by a single power-law function.
\subsection{Color Evolution}
\begin{figure}[htbp]
\center
\includegraphics[width=\textwidth]{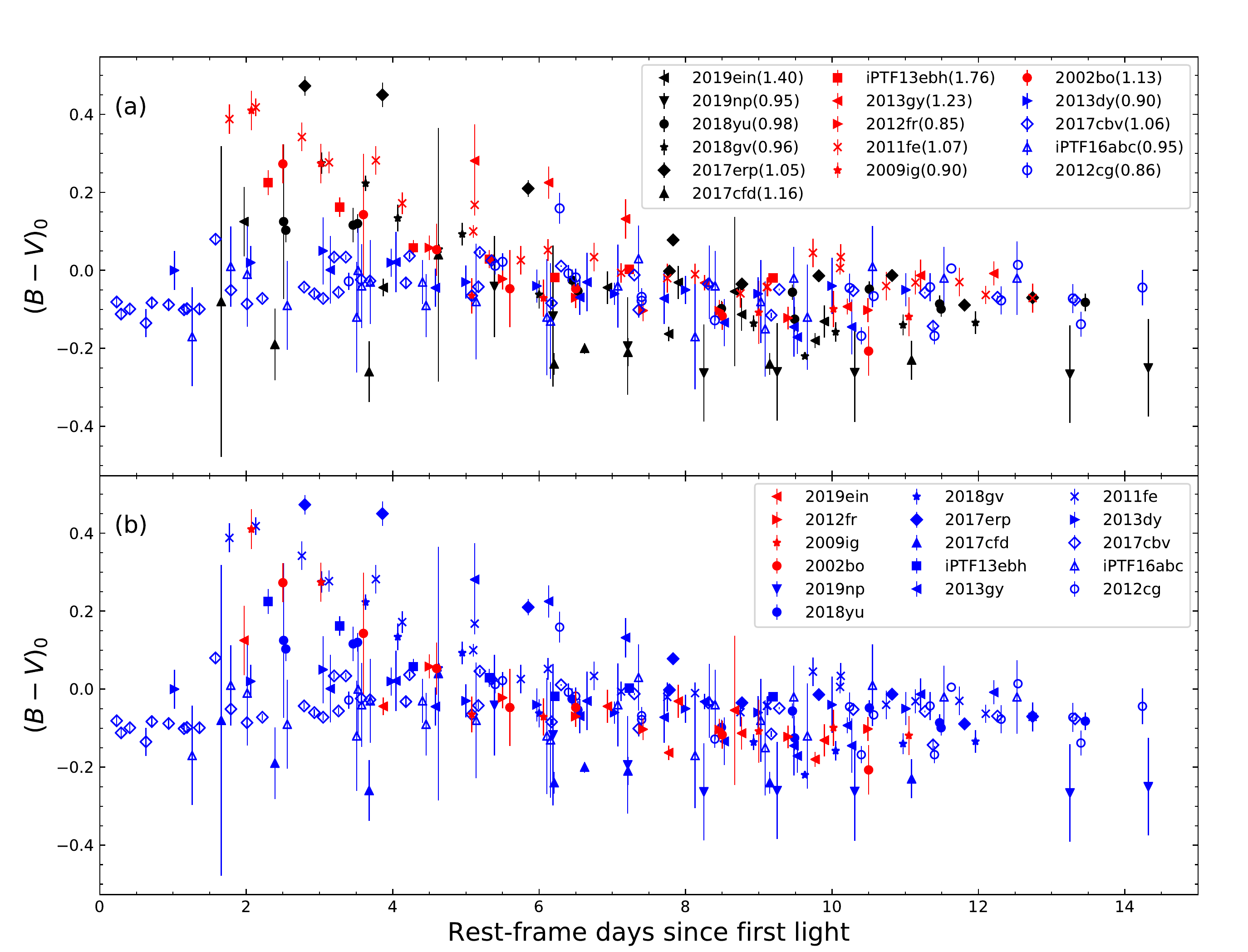}
%\vspace{-0.0cm}
\vspace{0.2cm}
\caption{($B-V$)$_0$ color curves of our sample. All curves have been corrected for both reddening and time dilation.       Each SN has the same symbol in two panels (but may have different colors). {\it Upper panel:} The SNe included in \cite{2018ApJ...864L..35S} are marked in red or blue while our new sample are in black. The $\Delta m_{15}$($B$) of each SN are listed in the bracket. {\it Lower panel:} HV group and NV group are marked in red or blue, respectively. Open symbols represent three 91T/99aa-like SNe Ia. }
\label{BV} \vspace{-0.0cm}
%\vspace{-0.5cm}
\end{figure}

%Very early light and color curves can provide information for the explosion mechanism of SNe Ia. 
We display extinction corrected ($B-V$)$_0$ color curves of our sample %before $-5$ days 
in Figure \ref{BV}. %Reddening corrections have been applied to all color curves. 
For those SNe with known extinction, the values given in the literature are used. Otherwise, we fit the multi-band light curves with SNooPy2 \citep{2011AJ....141...19B} to get the host galaxy extinction in addition to the Galactic extinction \citep{2011ApJ...737..103S}.      Intrinsic colors from \cite{2014ApJ...789...32B} are used to infer the host extinction $E(B-V)_{host}$ as well as $R_{Vhost}$. SN~2019ein is the only object with both new and published photometry, our new results are consistent with the previous works \citep{2020ApJ...893..143K,2020ApJ...897..159P}.

At four days after the first light, % the magnitude dispersion is diverse, $\sim$ 0.6 mag, and 
the ($B-V$)$_0$ color ranges from $\sim-$0.2 mag to 0.4 mag. This color dispersion gradually decreases to 0.3 mag several days later. Among our sample, SN~2017erp shows the reddest color in the early phase, i.e., ($B-V$)$_0$ $\sim$ 0.5 mag at $t\sim$ 2.8 days, and it quickly evolves to $\sim$ 0.05 mag at $t\sim$ 8 days, indistinguishable from other objects. SN~2017cfd shows the bluest color, i.e., $\sim-$0.2 mag at $t\sim$ 2.5 days, and the color mildly evolved after that. The distribution of ($B-V$)$_0$ seems to be continuous for our sample. \cite{2018ApJ...864L..35S} claim that the early-time colors of SNe Ia tend to show evidence for two populations, i.e., red and blue. 
We overplot their sample together with our new sample in Figure \ref{BV}(a).
%In our new sample, SNe 2017erp, 2018gv and 2019np  
Among the new sample, SN~2018yu and SN~2019ein seem to be located between the red and blue groups,      suggesting that the red and blue distinction is not sharp. It should be noted that the uncertainties are relatively large and fall within either the red or blue group for most of the data points in this sample. More early-time observations are needed to distinguish whether it is a continuous or bimodal distribution. %Therefore we propose the early colors are continuous, inconsistent with  \cite{2018ApJ...864L..35S}. 

To examine the color evolution for different sub-types of SNe Ia, we marked HV and NV group with different colors as shown in Figure \ref{BV}(b). One can see that the HV objects initially have red colors and then evolve bluewards with ($B-V$)$_0~\sim -$0.1 mag at $t\sim$ 5 days. The distribution for the NV group is much wider in the early phase, ranging from $-$0.2 mag to +0.4 mag. The 91T/99aa-like objects are hot and luminous, with some common features such as weak Si II features around the maximum and early excess flux \citep{2018ApJ...865..149J}. In the ($B-V$)$_0$ color evolution diagram, these slow decliners tend to dominate the blue side at the early time. %We use open symbols to represent three 91T/99aa-like objects, SNe 2012cg, 2017cbv and iPTF16abc, in Figure  \ref{BV}. All of the three SNe have blue colors and belong to NV group.
%SNe 2019ein, 2018yu and 2018gv are redder than 
\subsection{Carbon imprint}
To investigate the unburnt carbon in the ejecta, we examine the dominant C~II absorption feature in optical spectra (i.e., C~II~$\lambda$6580) of the spectra. This feature appears as a notch or a plateau on the red edge of Si~II~$\lambda$6355 absorption feature. 
%Figure \ref{carbon} displays the last detection (blue dots) or non-detection limit (yellow arrows) of C~II $\lambda$6580 features. 
We display the C~II~$\lambda$6580 region in the early spectra of all objects in Figure \ref{carbon}.
Objects with (without)  C~II $\lambda$6580 feature belong to NV (HV) group except for one HV object, SN~2019ein, which has a C~II~$\lambda$6580 feature that quickly disappears within three days of the first light.      This trend is consistent with \cite{2011ApJ...732...30P}. 
%The last detection of C~II  $\lambda$6580 ranges from 1 day to 21 days after first light. 
%The non-detection of carbon features among most HV objects is consistent with \cite{2014MNRAS.444.3258M}. 
The mean $\Delta m_{15}$($B$) of objects with carbon features (carbon-positive objects) and objects without carbon features (carbon-negative objects) are 1.08 $\pm$ 0.03 mag and 1.06 $\pm$ 0.03 mag, respectively, suggesting there is no trend between light curve widths and C~II features as reported by \cite{2014MNRAS.444.3258M} and \cite{2011ApJ...743...27T}. 

The carbon-negative objects have ($B-V$)$_0 >$ 0.1 mag at the very early phase and evolve rapidly blueward, %while the carbon-positive objects have much wider early color distribution. % (see the inset of Figure \ref{carbon}).
consistent with one-dimensional He-detonation models with thin helium shell (see Figure 6 of \cite{2019ApJ...873...84P}),  which have little unburnt carbon left after the explosion (but see Section \ref{dis}). Two NV objects, SN~2011fe and iPTF13ebh, also have possible C~I lines detection in their NIR spectra \citep{2013ApJ...766...72H,2015AA...578A...9H}. %SN~2014J is located near the border between NV and HV groups. 
However, \cite{2019ApJ...871..250H} questioned the identifications of these features because their simulations do not show significant neutral carbon. 

\begin{figure}[htbp]
\center
\includegraphics[width=\textwidth]{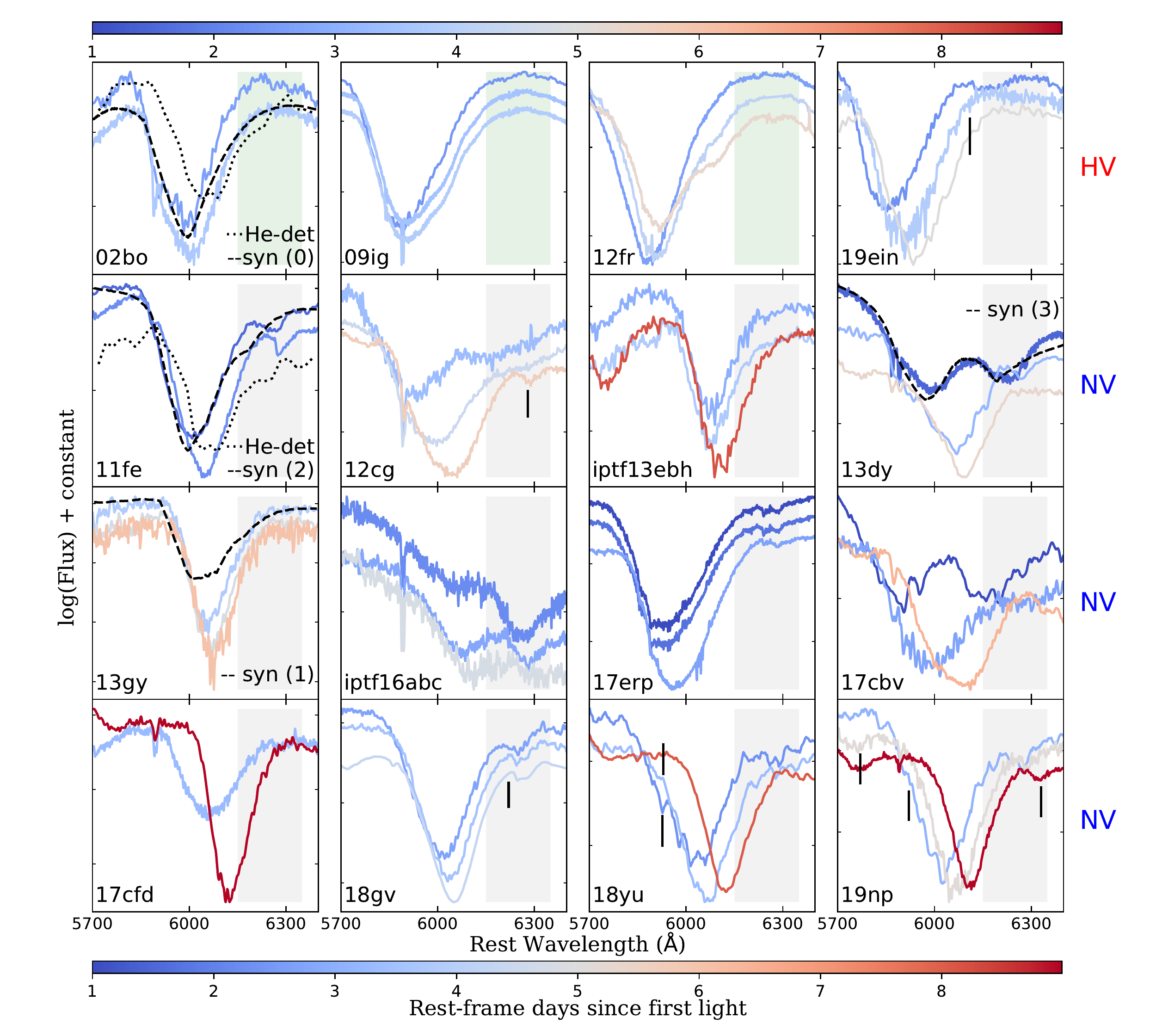}
%\vspace{-0.0cm}
\vspace{0.2cm}
\caption{C~II $\lambda$6580 detection in our sample. We display the first two or three spectra of each SN with the colors indicating the phases. The green and gray shadows show C~II~$\lambda$6580 regions for carbon-negative and carbon-positive SNe, respectively. The SNe in the first row are HV objects while the others are NV objects. We also plot the He-detonation (see Section \ref{dis}) model spectra (black dotted lines) synthesized at $t=$+5 days from the explosion in SN~2002bo ({\bf n$_1$} direction, see Figure \ref{fig:carbon}) and SN~2011fe ({\bf n$_3$} direction) panel,      as well as \texttt{SYNAPPS} fit to the observed spectra of SNe~2002bo, 2011fe,  SN~2013dy and SN~2013gy at $t\sim$+4 days with the $s_c$ value labeled in the legends (black dashed lines).      New spectra are labeled by vertical black lines.} 
\label{carbon} \vspace{-0.0cm}
%\vspace{-0.5cm}
\end{figure}
\iffalse
\begin{figure}[htbp]
\center
\includegraphics[width=\textwidth]{carbon.pdf}%plot.py
%\vspace{-0.0cm}
\vspace{0.2cm}
\caption{C~II $\lambda$6580 detection of our sample. We plot last detection times for carbon-positive objects (yellow dot) and last non-detection times for carbon-negative objects (black arrow). The blue dash line is $v~=$ 11,800 km~s$^{-1}$ which is the border between HV and NV group. The inset has the same data as Figure \ref{BV}, but with different colors and shapes. The yellow lines are carbon-positive ones while the black lines are carbon-negative ones.}
\label{carbon} \vspace{-0.0cm}
%\vspace{-0.5cm}
\end{figure}
\fi

\begin{figure}[htbp]
\center
\includegraphics[width=0.8\textwidth]{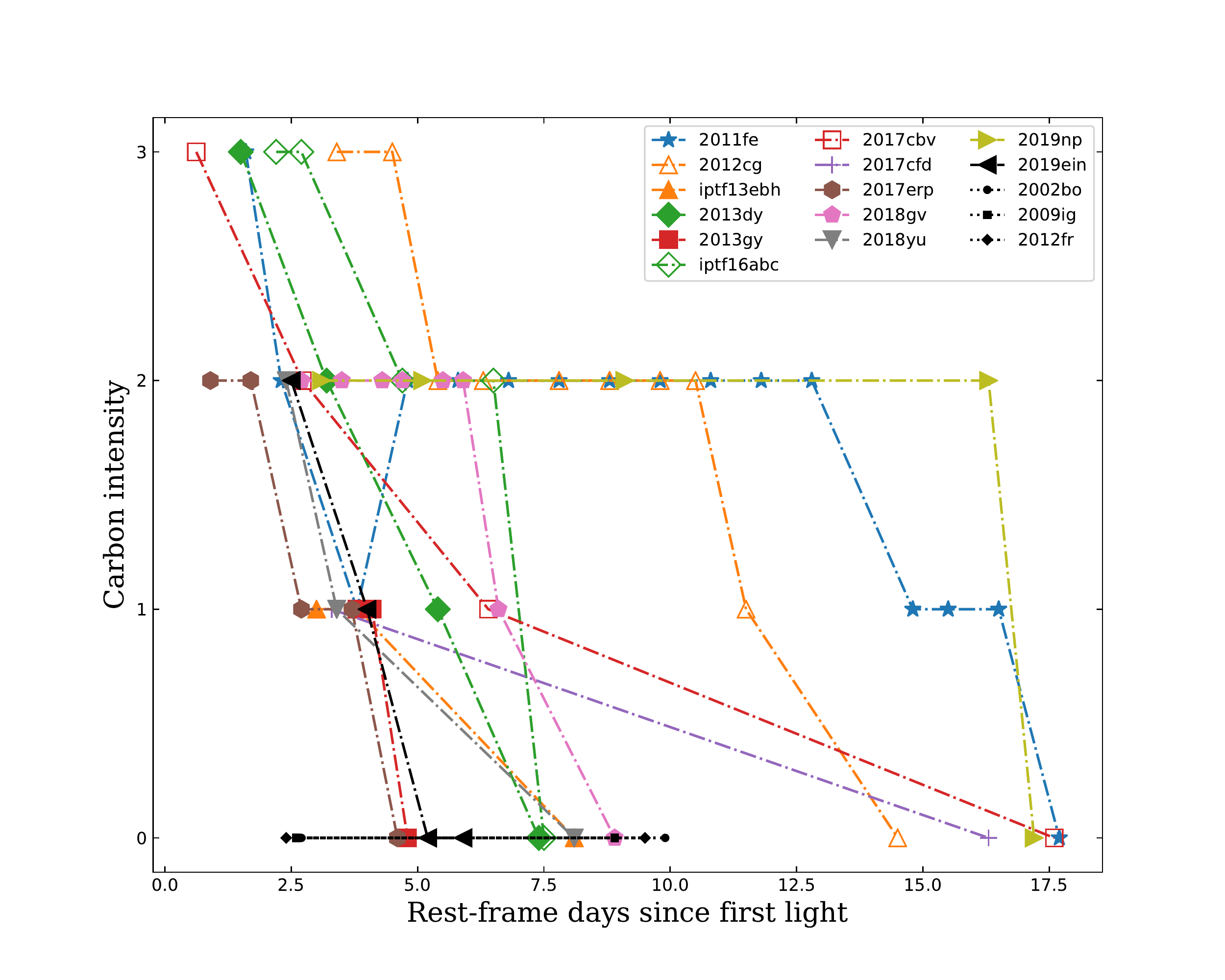}
%\vspace{-0.0cm}
\vspace{0.2cm}
\caption{The C~II $\lambda$6580 intensity in the spectra, evaluated for our sample. The HV objects are marked by black symbols while the NV objects by color filled symbols. The 91T/99aa-like objects are marked by open symbols.} 
\label{score} \vspace{-0.0cm}
%\vspace{-0.5cm}
\end{figure}

As the C~II~$\lambda$6580 features are usually weak and sometimes appear as a plateau, it is not easy to measure the pseudo-equivalent widths (pEWs). Therefore we use a semi-quantitative score ($s_c$ to describe their intensity. The scoring criteria and examples are as follows:
\begin{itemize}
\item 0: no C~II~$\lambda$6580 feature, such as the first spectrum of SN~2002bo;
\item 1: the feature appears as a plateau, such as the first spectrum of SN~2013gy;
\item 2: the feature appears as a notch, such as the first spectrum of SN~2011fe;
\item 3: the feature is strong enough to affect the profile of Si~II~$\lambda$6355 feature, such as the first spectrum of SN~2012cg.
\end{itemize}
We use the highly parameterized code \texttt{SYNAPPS} \citep{2011PASP..123..237T} to identify the carbon absorption features. The difference between 0 and 1 is obtained by comparing the spectra with carbon-free \texttt{SYNAPPS} fit, i.e., if the spectrum has no absorption feature but deviated from the fit, we score 1, otherwise we score 0. Other spectra are scored by eye inspection.      Four representative SYNAPPS fits for $s_c=0-3$ are overplotted in Figure \ref{carbon}. We plot the scores of all early spectra in Figure \ref{score}. This analysis confirms the trend mentioned above, namely that HV objects have quickly disappearing C~II~$\lambda$6580 feature or no such feature at all while NV objects generally have such a feature and last for a longer time. All three 91T/99aa-like objects also have strong C~II feature at early phase. We will discuss the implications of these results for the explosion mechanisms in Section \ref{dis}.
\subsection{Ca~II NIR triplet features} \label{sec:ca}
We use multiple Gaussian profiles to fit the photospheric-velocity features (PVFs) and HVFs of %several spectral lines including %Si~II $\lambda$6355,O~I $\lambda$7774 and 
Ca~II NIR triplet (CaIR3). The velocity and pEW of each absorption feature can be measured from the fit.
\iffalse
We display the pEW evolution of O~I $\lambda$7774 in Figure \ref{o1}. The O~I $\lambda$7774 features of the HV group are weak while the NV group have both weak and strong such feature. Oxygen is the ingredient of the progenitor and also the product of carbon burning, weaker oxygen lines indicate more complete burning which is consistent with non-detection of carbon signal among HV objects.

\begin{figure}[htbp]
\center
\includegraphics[width=\textwidth]{d_ewo.pdf}
%\vspace{-0.0cm}
\vspace{0.2cm}
\caption{pEW evolution of O~I $\lambda$7774. The circles represent the NV objects while the stars represent the HV objects.}
\label{o1} \vspace{-0.0cm}
%\vspace{-0.5cm}
\end{figure}
\fi
Previous studies indicate that the HVFs of CaIR3 are common in the spectra of SNe Ia \citep{2015ApJS..220...20Z}.      We plot the CaIR3 region in the early spectra of our sample in Figure \ref{fig:ca}.
In Figure \ref{ca2}, we investigate the velocity evolution of this spectral feature, and found that all HV objects tend to % because its spectra do not cover this range)
have very high velocity, i.e., $>$ 30,000 km s$^{-1}$, at $t\sim$ +3 days. In comparison, the CaIR3 HVFs of all NV objects %(no data for SN~2017cfd) 
tend to have velocities below 30,000 km s$^{-1}$, except two luminous 99aa-like objects, iPTF16abc and SN~2017cbv. Among the NV objects, SN~2017erp has the highest velocities with an early-time velocity of $\sim$ 29,000 km s$^{-1}$. The velocities of SN~2009ig and SN~2012fr are always above the NV groups while the velocity of SN~2019ein declines rapidly and becomes lower than the NV groups. The median velocity of the CaIR3 HVFs of HV group is $\sim$ 6,000 km~s$^{-1}$ higher than that of NV group at the early phase, and the difference gradually decreases to $\sim$ 2,000 km s$^{-1}$ at $\sim$ +14 days after explosion.

\begin{figure}[htbp]
\center
\includegraphics[width=\textwidth]{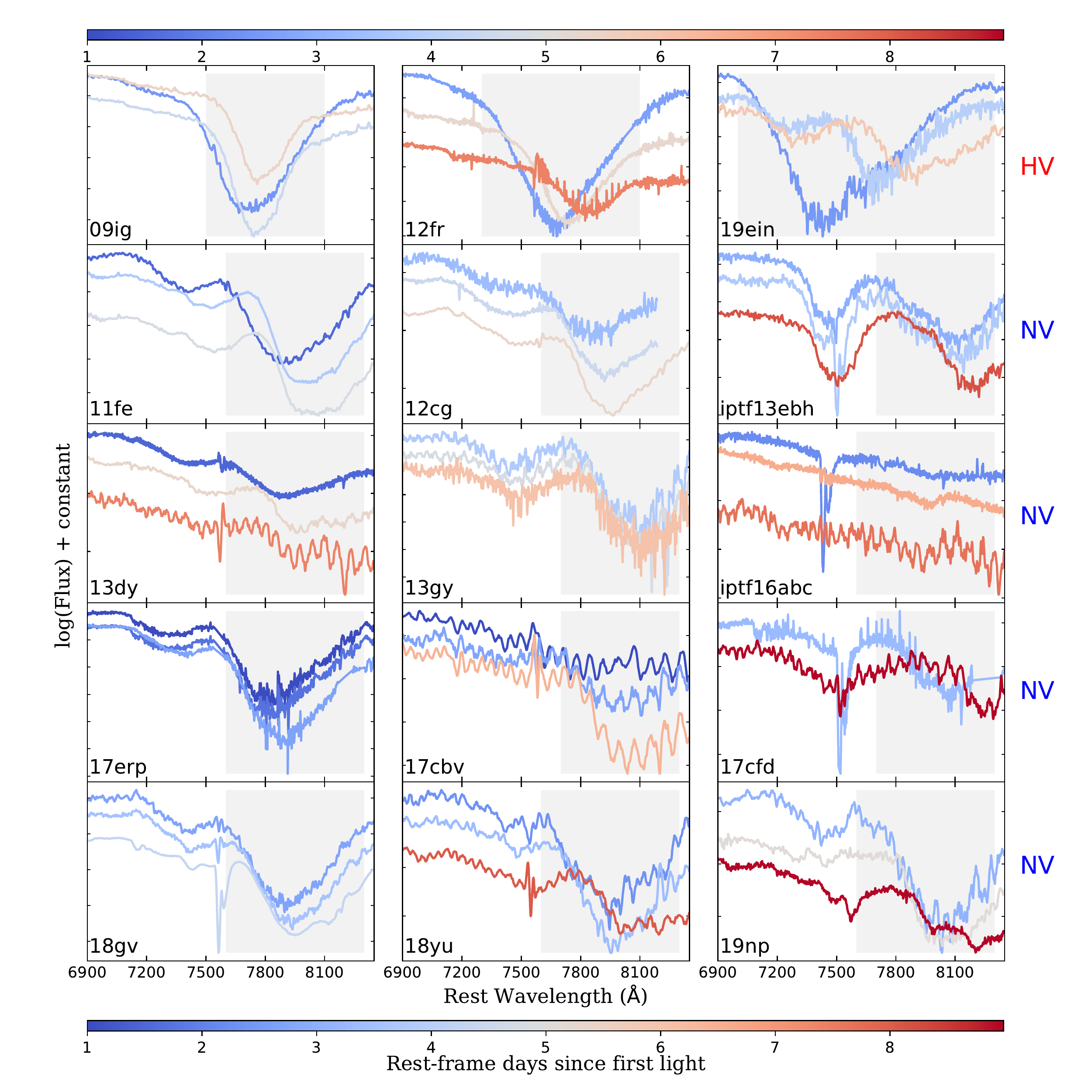}
%\vspace{-0.0cm}
\vspace{0.2cm}
\caption{Similar to Figure \ref{carbon}. We display the CaIR3 region for the first two or three spectra of each SN with the colors indicating the phases.} 
\label{fig:ca} \vspace{-0.0cm}
%\vspace{-0.5cm}
\end{figure}
%This difference could be explained with He-detonation scenario (see Section \ref{dis}). A correlation between evolution of these HVFs velocity and photospheric O~I $\lambda$7774 velocity among HV objects is also identified, implying the HVFs are originated from outer ejecta rather than CSM.
\begin{figure}[htbp]
\center
\includegraphics[width=\textwidth]{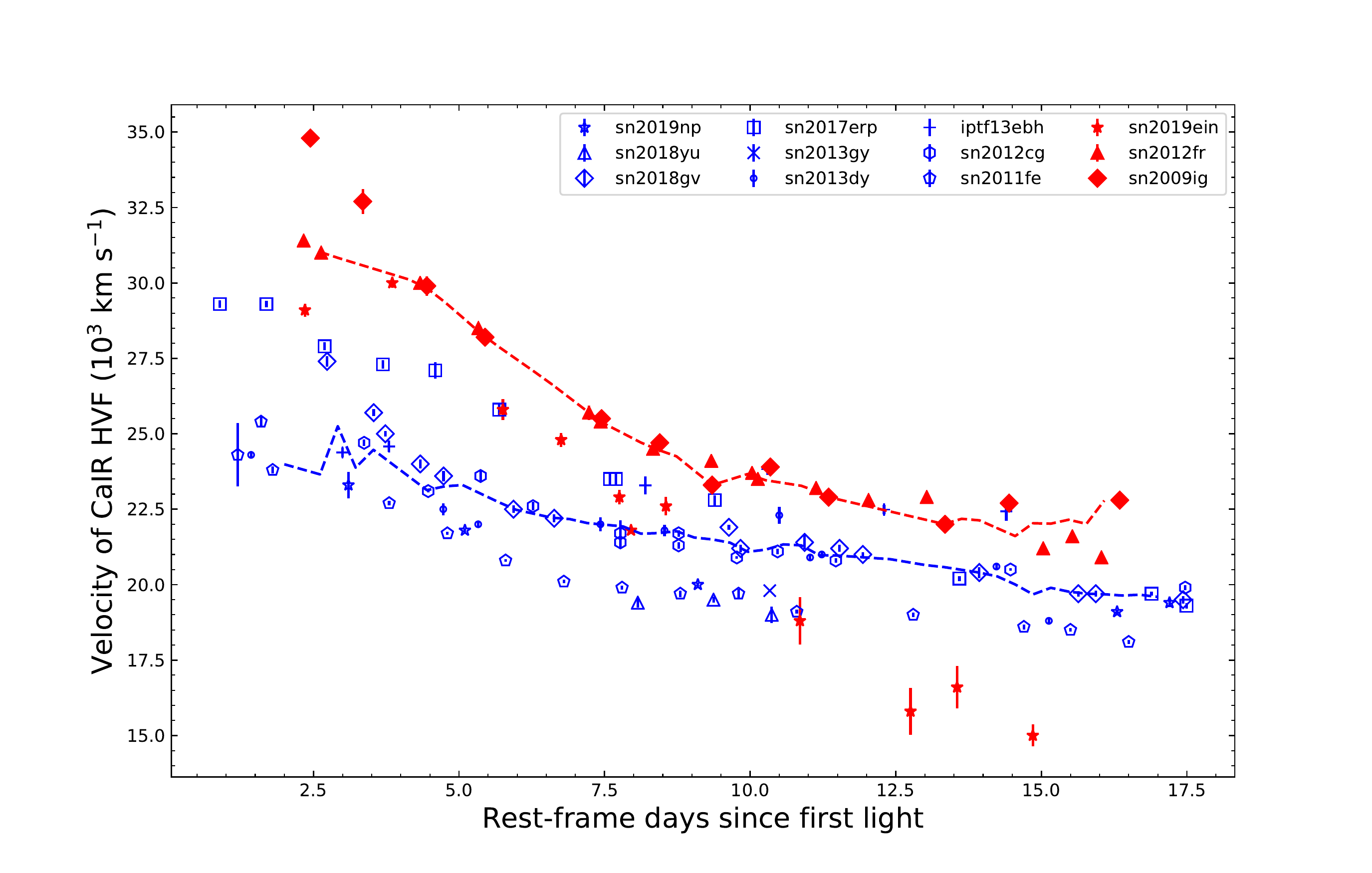}%spectra.py
%\vspace{-0.0cm}
\vspace{0.2cm}
\caption{Velocity evolution of CaIR3 triplet. The blue symbols represent the NV objects while the red symbols represent the HV objects. The dash lines are median curves for HV (red) and NV (blue) group.}
\label{ca2} \vspace{-0.0cm}
%\vspace{-0.5cm}
\end{figure}
\subsection{Host galaxy mass}
\begin{figure}[htbp]
\center
\includegraphics[width=\textwidth]{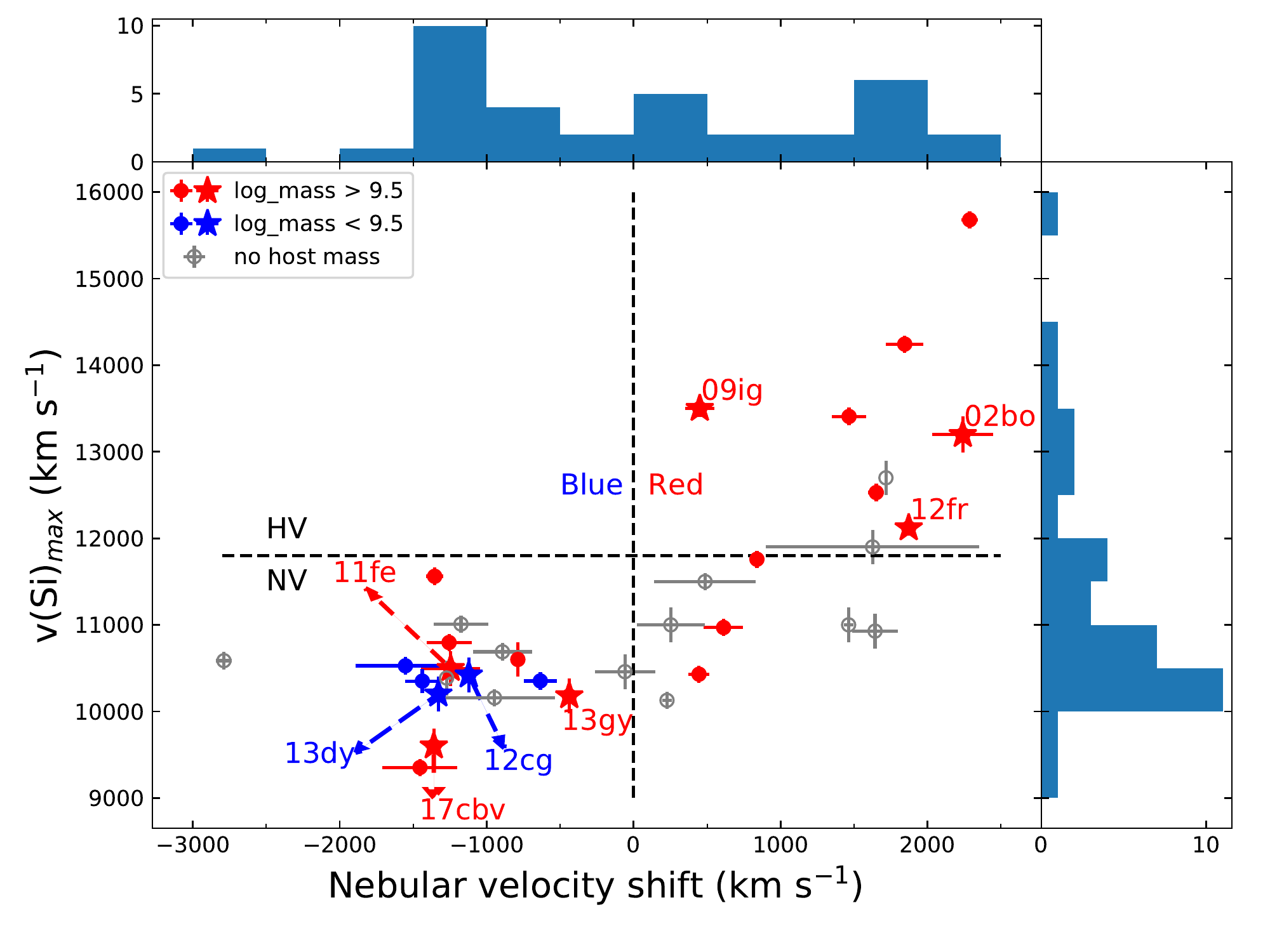}%spectra.py
%\vspace{-0.0cm}
\vspace{0.2cm}
\caption{Si~II velocity measured from near-maximum-light spectra versus the velocity shift of the [Fe II] 7155~\AA~feature inferred from the late-time spectra. The colors indicate the host galaxy masses \citep{2015MNRAS.446..354P,2018MNRAS.477.3567M,2019ApJ...874...32R,2020ApJ...895L...5P}.      The SNe in our sample are marked by stars and labeled by their names. The horizontal and vertical dash lines are the borders between HV/NV group ($v_{\rm Si,max}$ = 11,800 km~s$^{-1}$) and blue/redshift of the nebular velocities ($v_{\rm [Fe~II]} = 0$), respectively. We also plot histogram of $v_{\rm [Fe~II]}$ (top panel) and $v_{\rm Si,max}$ (right panel).}
\label{host} \vspace{-0.0cm}
%\vspace{-0.5cm}
\end{figure}

In order to constrain the physical origin of SNe Ia by combining early- and late-time observations and host galaxy properties, we study $v_{\rm Si,max}$, the velocity shift of late-time [Fe~II] $\lambda$7155 features, $v_{\rm [Fe~II]}$, and the host galaxy mass of the sample (see Figure \ref{host}). The spectral features from inner Fe-group regions dominate the late-time spectra, including [Fe~II] $\lambda$7155, which is the strongest optical feature formed from the deflagration ashes. The SNe Ia with higher $v_{\rm Si,max}$ have more redshifted $v_{\rm [Fe~II]}$, implying asymmetric explosions for some SNe Ia \citep{2010Natur.466...82M,2018MNRAS.477.3567M}. We add several SNe Ia with these parameters from the literature \citep{2015MNRAS.446..354P,2018MNRAS.477.3567M,2019ApJ...874...32R,2020ApJ...895L...5P}. We measure $v_{\rm [Fe~II]}$ of SNe~2017cbv and 2013gy using the Gaussian fit (see Section \ref{sec:ca}) and the host galaxy mass of SNe~2002dj, 2007le, 2010ev, 2013cs, 2013gy and 2019ein,  using the same method adopted in \cite{2020ApJ...895L...5P}.      \cite{2015MNRAS.446..354P} and \cite{2020ApJ...895L...5P} found that HV objects only occur in galaxies with $M_{stellar} >3 \times 10^9~(\sim 10^{9.5}$) $M_\odot$. Therefore we choose the boundary between massive and low-mass galaxies as log($M_{stellar}/M_\odot$) = 9.5. %In addition to the trend discovered before that 
We notice that all HV objects have redshifted $v_{\rm [Fe~II]}$ while NV objects have both redshifted and blueshifted $v_{\rm [Fe~II]}$. Moreover, all objects with redshifted $v_{\rm [Fe~II]}$ tend to reside in massive galaxies while objects with blueshifted $v_{\rm [Fe~II]}$ are located in both massive and low-mass galaxies. The $v_{\rm [Fe~II]}$ seems to have a double-peak distribution (see the top panel of Figure \ref{host}) with one at $\sim-$1000 km~s$^{-1}$ and the other at $\sim$+1500 km~s$^{-1}$, respectively, which may indicate two populations. Nevertheless, the Si II velocity distribution measured around the maximum light does not reproduce the double Gaussian profile as reported by \cite{2013Sci...340..170W}, 
which means that our sample size is still too small. According to the mass-metallicity relation \citep{2004ApJ...613..898T,2008ApJ...681.1183K}, massive galaxies have on average higher metallicity than low-mass galaxies, suggesting that objects with redshifted $v_{\rm [Fe~II]}$ (including all HV objects) have metal-rich progenitor environments. We will discuss the implications of these phenomena in Section \ref{dis}.
\section{discussion}\label{dis}
\subsection{     Early-time $(B-V)_0$ color distribution}
In this section, we further examine the differences and connections between NV and HV objects and provide implications for the explosion mechanism. \cite{2018ApJ...864L..35S} studied a sample of 13 SNe Ia, suggesting that they can be divided into red and blue groups according to the early $B-V$ color, and the red group has a rapid evolution from red to blue. \cite{2018ApJ...865..149J} reached a similar conclusion with 23 young SNe Ia. \cite{2020ApJ...892..142H} also found that these two groups may exist by adding a sample of six more supernovae, although we note that the distinction between these two sub-classes is less significant when these events are added (see Fig. 5 of \cite{2020ApJ...892..142H}). Our sample largely overlaps with the samples used in the above works, and the results inferred from the common sample are consistent. \cite{2020ApJ...902...48B} studied the $g-r$ (similar to $B-V$) curves of 65 SNe Ia from ZTF, discovered within five days from the first light, and they found that the early colors were uniformly distributed and did not show any evidence for two separate groups.      Our research adds a few new objects to this sample with early observations and also shows that the $B-V$ colors do not have a distinct bimodal distribution, either. One or more mechanisms, including companion interaction, $^{56}$Ni distribution and CSM interaction can affect the early-time light and color curves. However, quality of current color curves is not high enough to distinguish different processes, and it is difficult to  constrain the explosion mechanism from them.
\subsection{The link between He-detonation model and the early-phase features}
\begin{figure}[htbp]
\center
\includegraphics[width=1\textwidth]{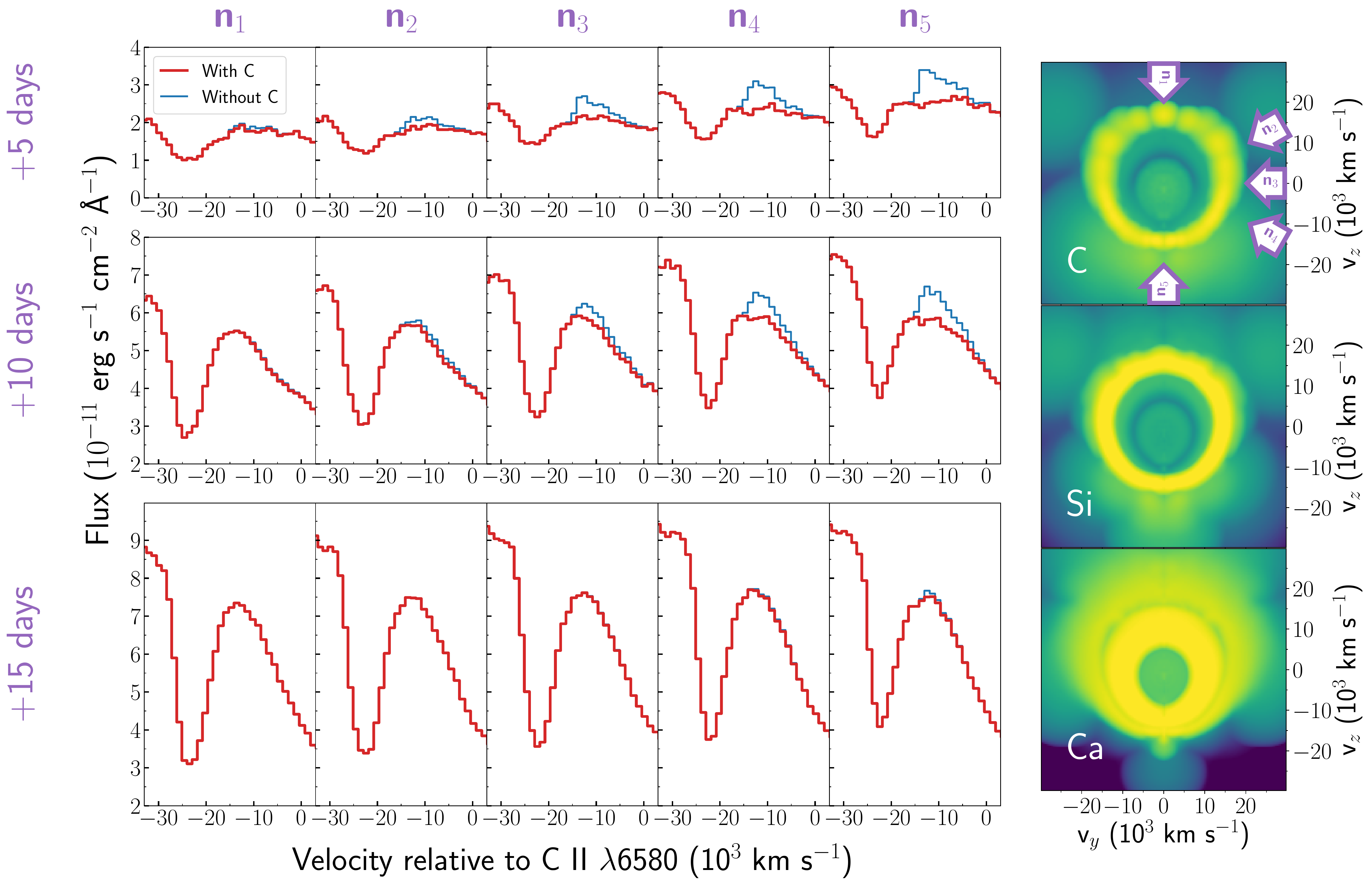}
\vspace{0.2cm}
\caption{Carbon features predicted with \texttt{ARTIS} \citep{2009MNRAS.398.1809K} for the double-detonation ``model~3m'' from \cite{2010ApJ...719.1067K}. {\it Left panels}: wavelength region around the Si~II~$\lambda$6355 and C~II~$\lambda$6580 profiles at three different epochs (5, 10 and 15 days since explosion from top to bottom) and for five different observers from North (left) to South (right). The predicted spectra are shown as a function of the C~II~$\lambda$6580 velocity in red, while blue lines are spectra where carbon has been artificially removed. Therefore, the difference between the two spectra (red and blue lines) in each panel highlights the carbon contribution to the spectral region. {\it Right panels}: carbon (top), silicon (middle) and calcium (bottom) mass distribution in an inner slice of the model ($yz$ plane), with the five observer viewing angles \textbf{n}$_1$-\textbf{n}$_5$ marked with arrows.      The maps are shown 7.8\,s after the explosion when the ejecta are already in homologous expansion, i.e. densities drop with time as $\rho\propto t^{-3}$ but the C, Si and Ca distributions are unchanged at the epochs investigated (+5, +10, +15 days).}
\label{fig:carbon} \vspace{-0.0cm}
\end{figure}

The absence or presence of very weak carbon signal among spectra of HV group requires an explosion mechanism with high carbon-burning efficiency. The He-detonation scenario leaves almost no carbon after the explosion \citep{2019ApJ...873...84P,2019ApJ...878L..38T}. However, small amount of carbon could leave weak orientation-dependent absorption features in the spectra, which are hardly observable at the north side\footnote{We regard the helium detonation point (\textbf{n}$_1$ direction in Figure \ref{fig:carbon}) as north pole throughout the paper.}  (\textbf{n}$_1$ - \textbf{n}$_3$ in Figure \ref{fig:carbon}) but could be strongest at the south pole (\textbf{n}$_5$ in Figure \ref{fig:carbon}). Therefore,      a portion of NV objects with carbon absorption features could be observed from the south side of He-detonation while HV objects could be observed from the north side. This is consistent with the viewing-angle effect of Si II velocity (around maximum) in the He-detonation model, where ejecta move at higher velocities in the northern hemisphere and lower velocities in the southern hemisphere \citep{2016MNRAS.462.1039B,2019ApJ...878L..38T}. In Figure~\ref{fig:carbon}, we show the spectral region around the Si~II~$\lambda$6355 line predicted for the He-detonation ``model~3m'' presented by \cite{2010ApJ...719.1067K}. This model is a modified version of ``model~3'' from \cite{2010AA...514A..53F} where the helium shell is polluted by 34$\%$ of $^{12}$C to reduce the production of iron-group elements in the shell and thus provide a better match to the observed spectra of SN Ia (see \citealt{2010ApJ...719.1067K} for more details). Simulations have been carried out using the radiative transfer code \texttt{ARTIS} \citep{2009MNRAS.398.1809K}. 

Spectra produced by our He-detonation model are shown in Figure~\ref{fig:carbon} for five different viewing angles from north to south and at three different epochs: 5, 10 and 15 days after the explosion.
Similarly to the observations, the C~II~$\lambda$6580 feature in the model spectra appears as a notch or a plateau on the red edge of Si~II~$\lambda$6355 absorption feature. This carbon feature is absent for observers at the north pole and becomes stronger when moving towards the southern hemisphere, consistent with the idea that HV objects with little or no carbon could be He-detonation explosions viewed from the ignition side, while      a portion of NV objects are viewed from the opposite side. Figure~\ref{fig:carbon} also shows how the presence of C~II~$\lambda$6580 quickly weakens with time and disappears 15 days after the explosion (corresponding to $\sim$2 to 5 days before peak going from south to north) even for an observer in the southern hemisphere. We plot the He-detonation spectra at 5 days from two directions, \textbf{n}$_1$ and \textbf{n}$_3$, in SN~2002bo and SN~2011fe sub-figure of Figure \ref{carbon}. Despite the line profiles of Si~II~$\lambda$6355 do not entirely match the observations, the non-detection of C~II feature in the spectrum of SN~2002bo is reproduced while SN~2011fe does have such a feature in the early spectra. The DDT model is another candidate mechanism which leaves unburned carbon near the surface due to the deflagration process \citep{2013MNRAS.429.1156S}. Therefore it cannot explain the absence of carbon in the outer ejecta of HV objects.%oafThe weak O~I clines in HV objects and some NV objects are consistent with rather complete burning. 

The HVFs have three proposed origins: density enhancement (DE), abundance enhancement (AE) and ionization effect (IE) \citep{2005ApJ...623L..37M,2008ApJ...677..448T}. The extremely high velocities of CaIR3 is ubiquitous among the early spectra of HV objects, implying that they may originate from the same mechanism. Synthesized Ca from He-detonation at the north side can reach a velocity of more than 30,000 km~s$^{-1}$ while the Ca at the south side has much lower velocity (Figure \ref{fig:carbon}).
\cite{2019ApJ...878L..38T} also argue that the double-layer structure in the He-detonation model may produce AE in the outer ejecta and thus explain both HVFs and PVFs of Si and Ca. Interaction between SN ejecta and CSM can produce DE or IE effect \citep{2018MNRAS.476.1299M,2019MNRAS.484.4785M}, though it is difficult to consistently demonstrate the difference in HVFs of HV and NV SNe Ia. The DDT simulations that is most consistent with the observations of normal SNe Ia have multiple ignition sparks (N40 and N100 in \cite{2013MNRAS.436..333S}, which has 40 and 100 sparks, respectively), therefore they are close to spherical symmetry.       In principle, the DDT model can result in high-velocity ejecta (e.g., \cite{1991AA...245..114K}). Although some 3D simulations (e.g., the N1 and N3 models from \cite{2013MNRAS.429.1156S}) can reproduce the high-velocity intermediate mass elements (IME) at the outermost layer of the supernova ejecta, the $^{56}$Ni yields seems to be higher than that observed for normal SNe Ia. Current DDT models seem to have difficulties in producing the HVF of some SNe Ia, especially those extremely HVFs seen in SN~2019ein \citep{2020ApJ...897..159P}.
%is do not leave HVFs signals in the simulated spectra \citep{2013MNRAS.436..333S}.} 
%Another evidence from our analysis is the extreme HVFs of CaIR3 in outer ejecta during the first few days of HV objects which could originate from 
%\subsection{Implications on the physical origins of SNe Ia from maximum and nebular phase}
\subsection{Bimodal Distribution of Si~II and [Fe~II] Velocities}
HV and NV groups were initially classified according to the $v_{\rm Si,max}$ (i.e., \cite{2009ApJ...699L.139W}). 
%However, there may not be a sharp boundary between these two subtypes.
The observed $v_{\rm Si,max}$ distribution of SNe Ia tends to exhibit a double Gaussian distribution, while these two components have an overlapping regions (see Figure 1 of  \cite{2013Sci...340..170W} and Figure 3 of \cite{2020MNRAS.tmp.2978Z}). %According to these two profiles,
The HV profile is much broader, reaching a velocity as low as $<$ 10,000 km~s$^{-1}$, while the high-velocity tail of the NV profile extends a little into the HV region. If asymmetric He-detonation is the explosion mechanism responsible for the HV SNe Ia, %the velocity distribution within HV group could be explained as a geometrical effect, and
a portion of NV objects may have the same explosion mechanism but with different viewing angles, which can explain the higher-velocity and broader Gaussian distribution. Objects like SN~2017erp with several characteristics of HV group such as red early-time color, quickly disappeared C~II feature and relatively high velocity of CaIR3 HVFs, may have the same origin as the HV group, and their low $v_{\rm Si,max}$ might result from the geometrical effect. 

\cite{2010Natur.466...82M} and \cite{2018MNRAS.477.3567M} found that SNe Ia with higher $v_{\rm Si,max}$ tend to exhibit redshifted [Fe~II] features in the nebular phase, implying an asymmetric explosion mechanism for some SNe Ia. \cite{2015MNRAS.446..354P} and \cite{2020ApJ...895L...5P} found that HV objects tend to occur in massive host galaxies, while NV objects are located in both low-mass and massive host galaxies. Our finding that all HV objects have redshifted $v_{\rm [Fe~II]}$ while the NV counterparts have both redshifted and blueshifted $v_{\rm [Fe~II]}$ can fit into the sub-M$_{ch}$ He-detonation models, where one would observe a HV object with redshifted Fe-group features when viewing from He-detonation side (\textbf{n}$_1$ direction of Figure \ref{fig:carbon}),
and a NV object in the opposite side (south side) with blueshifted (or less redshifted) $v_{\rm [Fe~II]}$. The orientation-dependent Si-velocity trend is also in agreement with the synthetic spectra \citep[see Figure 4 of ][]{2019ApJ...878L..38T}. We also find that objects with redshifted $v_{\rm [Fe~II]}$, including all HV objects, are located in massive host galaxies (log($M_{stellar}/M_\odot$) $>$ 9.5), which implies that they have metal-rich progenitor environments. Metallicity plays an important role in the evolution of SNe Ia.  \cite{2020MNRAS.491.2902F} found that the sub-M$_{ch}$ explosion simulations of $\sim$Z$_\odot$ progenitors can explain the Ni/Fe abundance measurements for the majority of normal SNe Ia while only a small fraction (11\%) of their sample are in agreement with M$_{Ch}$ DDT explosion models. % If HV and some NV objects have sub-M$_{ch}$ progenitors, the high metallicities are consistent with their massive host galaxies. 
The sub-M$_{ch}$ models include both He-detonation and violent merger scenarios.  However, polarisation measurements disfavor the strongly asymmetric violent merger scenario while support He-detonation and DDT models which are more symmetric \citep{2016MNRAS.455.1060B,2016MNRAS.462.1039B}. Therefore, we regard these results as supports of He-detonation scenario. Metal-rich environments and progenitors are more likely to produce CSM around the progenitor systems, consistent with the observations that SNe Ia of HV group are more likely to have abundant CSM than the NV group. \citep{2012ApJ...752..101F,2012ApJ...749..126W,2019ApJ...882..120W}.
 %Other objects with normal velocity is inconsistent with this interpretation and might have different explosion mechanism.

\cite{2019ApJ...873...84P} claim that the explosions of WDs with different masses, rather than the different viewing angles, may cause the range of $v_{\rm Si,max}$. \cite{2015MNRAS.446..354P} and \cite{2020ApJ...895L...5P} confirmed that the HV group tends to have metal-rich progenitors. At a given mass, stars with higher metallicities generally produce less massive WDs, thus the HV group may originate from the sub-M$_{Ch}$ progenitors. However, the one-dimensional models studied by \cite{2019ApJ...873...84P} cannot explain the asymmetries revealed from the nebular observations. The CaIR3 features of these models are also too weak to account for the observations. 
\subsection{Possible Progenitor Systems for He-detonation Scenario}
To form a CO WD with a thin He shell as required by He-detonation model, we need the CO WD to accrete and accumulate helium from the companion star, either a He-star or a He-rich WD \citep{2017MNRAS.472.1593W,2018RAA....18...49W}. One of the main differences between these two scenarios is the delay times -- the time interval from the formation time of the primordial binaries to the supernova explosion. The CO WD + He-star channel has shorter delay times, therefore it should originate from younger populations \citep{2009MNRAS.395..847W,2017A&A...606A.136L}. \cite{2013Sci...340..170W} found that the HV objects, similar to Ibc supernovae, are located in the inner and brighter regions of host galaxies than the NV objects, which suggests a origin of younger progenitor populations. \cite{2020ApJ...895L...5P} found that the host galaxies of the HV group does not tend to be younger than the NV group by analyzing the global specific star formation of the host galaxies. However, the global properties may not reflect the local properties which was utilised by \cite{2013Sci...340..170W}. SN~2012Z, the first SN Ia with detection of the progenitor system in prediscovery images, is believed to have a He-star as the companion star \citep{2014Natur.512...54M}. V445 Pup, the only He nova discovered so far, is a WD with a $\sim$$M\rm_{Ch}$ mass and can accumulate accreted He from its He-star companion after the outburst \citep{2008ApJ...684.1366K,2009ApJ...706..738W}. 
%making it a strong candidate for SN Ia  \citep{2018RAA....18...49W}. 
It should be noted that a WD with a $\sim$$M\rm_{Ch}$ mass may not produce normal SN Ia through He-detonation explosion because the central density is too large      and produce too much $^{56}$Ni \citep{2010AA...514A..53F}. A system similar to V445 Pup but with a less massive CO WD can be a better candidate for He-detonation explosion. Furthermore, the ejecta of the He nova outbursts may form circumstellar dust, which are more common around HV objects \citep{2019ApJ...882..120W}. Combining the younger progenitor environments, prediscovery progenitor system detection, possible progenitor candidate and CSM around progenitor system, we tend to favor CO WD + He-star system for the progenitor system of He-detonation scenario, although the possibility of CO WD + He-rich WD channel cannot be fully excluded.%liu's candidates of He WD, birth rate
\section{Conclusion}\label{con}
We analyze a sample of 16 SNe Ia with early spectral and photometric observations, including four HV objects: SNe~2002bo, 2009ig, 2012fr and 2019ein. 
%SN~2014J is located at the boundary of HV and NV, and the first observation is late. 
We investigate the $B-V$ color evolution of the sample and find that the color distribution tends to be continuous. %, i.e., there is no apparent bimodal distribution. 
The color dispersion is $\sim 0.6$ mag at $t=2\sim3$ days since explosion and gradually decreases to $\sim 0.3$ mag at $\sim$ two weeks after that. In the early spectra of SNe Ia, HV objects do not have or have weak C~II $\lambda$6580 feature while this feature is prominent in NV objects. This contrast may be understood in terms of He-detonation model, where the HV objects are observed from the the helium detonation side while NV objects are observed from the opposite side. Moreover, the CaIR3 HVFs of HV objects are apparently higher than those of NV objects, exceeding 30,000 km~s$^{-1}$ in very early phase, and these HVFs can be also naturally explained by observing the SNe Ia towards the helium detonation side. %model can naturally produce AE effect of Si and Ca in the northern hemisphere (     n}$_1$-     n}$_3$ in Figure \ref{fig:carbon}) through the first He detonation, and produce HVFs with extremely high velocities observed in the spectra. 
We analyze the relation between $v_{\rm Si,max}$, $v_{\rm [Fe~II]}$ and the host galaxy mass. All HV objects have redshifted $v_{\rm [Fe~II]}$ and are located in massive galaxies; while NV objects have both blue- and redshifted $v_{\rm [Fe~II]}$, and are located in both massive and low-mass galaxies. SNe with redshifted $v_{\rm [Fe~II]}$ require an asymmetric explosion model such as He-detonation model. The fact that all redshifted SNe are located in massive galaxies implies that they may have similar progenitor metallicities and explosion mechanisms.

We conclude that He-detonation model can simultaneously explain several characteristics of HV SNe Ia and a portion of NV objects, and the observed differences between them may be caused by different observation orientations. The presence of two populations is also favored by the Si II velocity distribution. %The overlapping double Gaussian distribution of  $v_{\rm Si,max}$ implies two population, one of which includes almost all HV objects and some NV objects. The velocity difference within this population can be explained by different viewing angles, which is consistent with He-detonation model. 
The current DDT models cannot explain the carbon signal and HVFs in the spectra. For He-detonation model, we propose that the CO WD + He-star progenitor system is more likely, though the CO WD + He-rich WD progenitor system cannot be fully excluded.
However, some features predicted by He-detonation simulation deviate from the observations, such as the rapidly decreasing light curve \citep{2018ApJ...854...52S,2019ApJ...878L..38T}. %The DDT scenario is still a possible explosion mechanism for some SNe, such as SN~2014J \citep{2019ApJ...882...30L}.
Whether HV and NV objects are intrinsically different or HV and a portion of NV objects have a common origin but different viewing angles requires more evidence. %The CSM around HV objects \citep{2012ApJ...752..101F,2019ApJ...882..120W} cannot be incorporated into He-detonation model. 
Nevertheless, our work confirms that HV group is a distinctive subgroup of SNe Ia. Future multi-dimensional and more realistic simulations may better match the observations. Larger sample of infancy SNe Ia, especially spectral observations, will provide more clues to the progenitor problem through color evolution, Fe-group elements distribution, HVFs features, etc. Combining early and nebular observations, as well as host galaxy properties, will help to better understand the physics underlying the observed subtypes of SNe Ia.

\acknowledgments
This work is supported by the National Natural Science Foundation of China (NSFC grants, 11633002, and 11761141001)). The authors gratefully acknowledge the Gauss Centre for Supercomputing (GCS) for providing computing time through the John von Neumann Institute for Computing (NIC) on the GCS share of the supercomputer JUQUEEN \citep{Stephan2015} at J\"ulich Supercomputing Centre (JSC). GCS is the alliance of the three national supercomputing centres HLRS (Universit\"at Stuttgart), JSC (Forschungszentrum J\"ulich), and LRZ (Bayerische Akademie der Wissenschaften), funded by the German Federal Ministry of Education and Research (BMBF) and the German State Ministries for Research of Baden-W\"urttemberg (MWK), Bayern (StMWFK) and Nordrhein-Westfalen (MIWF). J.Z. is supported by the National Natural Science Foundation of China (NSFC, grants 11773067, 11403096), by the Youth Innovation Promotion Association of the CAS (grant 2018081), and by the and Ten Thousand Talents Program of Yunnan for Top-notch Young Talents. L. Wang is supported by NSF award AST-1817099. 

We thank Y. Yang, S. Holmbo, P. Brown, W. Zheng, J. Vinko and K. Maguire for sharing the data used in this paper. We thank D. Townsley and J. Sun for helpful discussion on this paper.
\software{SWARP\ \ \citep{2002ASPC..281..228B},\ \ SExtractor\ \ \citep{1996AAS..117..393B},\ \ SCAMP\ \ \citep{2006ASPC..351..112B},\ \ Matplotlib\ \ \citep{hunter},\ \ NumPy\ \ \citep{numpy},\ \ SciPy\ \ \citep{scipy}, \ \ SNooPy2\ \ \citep{2011AJ....141...19B},  \ \ Lmfit  \ \ \citep{2016ascl.soft06014N},  \ \  \texttt{ARTIS} \ \ \citep{2009MNRAS.398.1809K,2015MNRAS.450..967B} and \ \ \texttt{SYNAPPS}\ \ \citep{2011PASP..123..237T}.}
{}

\end{document}